\documentclass[11pt]{article}
\usepackage{acl}

\usepackage{times}
\usepackage{latexsym}

\usepackage{amsmath}
\usepackage{amsfonts}
\usepackage[linesnumbered,ruled,vlined]{algorithm2e}
\usepackage{graphicx} 
\usepackage{booktabs}
\usepackage{array}
\usepackage{enumitem}
\usepackage{amsthm}
\usepackage{algpseudocode}

\usepackage[switch]{lineno}
\usepackage[utf8]{inputenc}

\usepackage{eqparbox}
\usepackage[nopar]{lipsum}
\usepackage{multirow}
\usepackage{makecell}

\usepackage{xcolor}

\usepackage{scrextend}
\deffootnote[.25in]{.25in}{.15in}{\makebox[.25in][r]{\thefootnotemark .\hspace{.15in}}}

\makeatletter

\newtheorem*{proof*}{Proof}

\usepackage{listings}
\usepackage{color}
\definecolor{codegreen}{rgb}{0.3,0.5,0.0}

\def\@fnsymbol#1{\ensuremath{\ifcase#1\or \dagger\or *\or \ddagger\or
   \mathsection\or \mathparagraph\or \|\or **\or \dagger\dagger
   \or \ddagger\ddagger \else\@ctrerr\fi}}
   
\title{GateFormer: Speeding Up News Feed Recommendation with Input Gated Transformers}

\author{Peitian Zhang \\
  Microsoft Research Asia \\
  Renmin University of China \\
  \texttt{namespace.pt@gmail.com} \\\And
  Zheng Liu \\
  Microsoft Research Asia \\
  \texttt{zhengliu@microsoft.com} \\}

\begin{document}
\maketitle

\begin{abstract}
News feed recommendation is an important web service. In recent years, pre-trained language models (PLMs) have been intensively applied to improve the recommendation quality. However, the utilization of these deep models is limited in many aspects, such as lack of explainability and being incompatible with the existing inverted index systems. Above all, the PLMs based recommenders are inefficient, as the encoding of user-side information will take huge computation costs. Although the computation can be accelerated with efficient transformers or distilled PLMs, it is still not enough to make timely recommendations for the active users, who are associated with super long news browsing histories.

In this work, we tackle the efficient news recommendation problem from a distinctive perspective. Instead of relying on the entire input (i.e., the collection of news articles a user ever browsed), we argue that the user's interest can be fully captured merely with those representative keywords. Motivated by this, we propose GateFormer, where the input data is gated before feeding into transformers. The gating module is made \textit{personalized}, \textit{lightweight} and \textit{end-to-end learnable}, such that it may perform accurate and efficient filtering of informative user input. GateFormer achieves highly impressive performances in experiments, where it notably outperforms the existing acceleration approaches in both accuracy and efficiency. We also surprisingly find that even with over 10-fold compression of the original input, GateFormer is still able to maintain on-par performances with the SOTA methods. 
\end{abstract}

\section{Introduction}
Online news platforms have become people's major access to real-time information. Nowadays, there have been huge amounts of daily news articles published by different sources, making it imperative to recommend users with personalized content automatically. Thanks to the development of deep learning techniques, deep neural networks, especially the pre-trained language models (PLMs), e.g., BERT \cite{Devlin2019BERT} and RoBERTa \cite{Liu2019Roberta}, are intensively utilized for high-quality news recommendation~\cite{wu_newsPLM,NewsBert}. Despite the improvement of recommendation quality, the PLMs based news recommenders are severely limited in efficiency. This is because the recommenders need to encode users' historical news browsing behaviors to capture their underlying interests. However, many real-world users are associated with super long histories, which will incur considerable computation costs when PLMs are employed as the encoding backbones. Although accelerations can be made with recent approaches, like distilled lightweight PLMs \cite{sanh2019distilbert} and efficient transformers \cite{tay2020efficient}, the computation cost will still grow dramatically with the input size. As a result, it remains a great challenge to make timely recommendations for active users.


In this work, we tackle the efficient news recommendation problem from a distinctive perspective. We argue that making news recommendations is more of capturing a user's underlying interests towards ``high-level semantics'', such as topics and entities, rather than memorizing what was exactly browsed by the user. Therefore, instead of relying on the entire content of the historically browsed news articles, the user's underlying interest is likely to be well represented with a small number of keywords. Driven by this motivation, we propose a novel efficient framework termed as \textbf{GateFormer}, where the high-informative fraction of user-side input can be filtered for making news recommendations. In GateFormer, a user is represented with two steps. Firstly, each input news article is processed by the gating module, which selects the top-$K$ informative keywords from the news text. Secondly, the filtered keywords from the entire user history are concatenated and encoded by transformers, generating the user representation. By setting $K$ to a reasonably small number, the encoding cost will be substantially reduced, as the user-side input becomes significantly smaller than its original size.

The gating module is highlighted by the following features. 1) The gating module is designed to be \textit{personalized}. While filtering the user-side input, the gating module estimates each word's importance w.r.t. user's underlying interest; if one word strongly correlates with the user interest, it will be selected by the gating module with a high priority. In this way, we are able to maximally preserve the keywords that truly matter to news recommendation. 2) The gating module is designed to be \textit{lightweight}. Knowing that the gating module needs to inspect the entire input, we implement it with tiny network components, like single-layer CNNs or GRU. By doing so, the extra time cost from the gating operation becomes almost ignorable to the computation cost incurred by transformers. 3) Considering that there is no supervision data for input filtering, the gating module is designed to be a differentiable component cascaded with transformer module; thus, it can be \textit{end-to-end learned} to select proper keywords for the optimal recommendation performance. 

Note that the filtered user input may naturally support inverted index (it can be used as an enhanced BM25 system) and provide keyword-based explanations for recommendation results, which are the additional benefits of using GateFormer.

To summarize, the major contributions of this work are listed as the following points.
\begin{itemize}
    \item We propose a novel efficient news recommendation framework called GateFormer. To the best of our knowledge, this is the first work that introduces a gating mechanism to accelerate the recommendation process. 
    \item Our gating module is designed to be personalized, lightweight, and end-to-end learnable, which enables representative keywords to be accurately filtered with little cost.
    \item Our experimental studies verify GraphFormers' effectiveness: it notably outperforms the existing acceleration approaches in both accuracy and efficiency; and it achieves on-par performances as the SOTA brute-force methods with a significantly compressed input.
\end{itemize}

\section{Related Work}
News recommenders aim to identify users' interested news articles based on their news reading histories \cite{wu2020mind,li2010contextual,das2007google}. In recent years, deep neural networks have been intensively applied to better represent the underlying semantics about news content and user behaviors. For example, in \cite{wang2018dkn}, entity embeddings are jointly learned together with user models, so as to introduce knowledge graph information into news recommendation; in \cite{wu2019neural}, the hierarchical attention networks (HAN) \cite{yang2016hierarchical} is adapted for the multi-view representation of news articles \cite{wu2019neural}; and in \cite{an2019neural}, long short-term memory networks are leveraged to capture users' time-sensitive interests. The latest news recommenders are built upon pre-trained language models \cite{wu2021newsbert,xiao2021training}, where PLMs like BERT \cite{Devlin2019BERT} and RoBERTa \cite{Liu2019Roberta} are employed as the backbones of news encoder. With the adoption of such highly expressive models, rich-semantic  embeddings can be generated, which substantially benefits the recommendation quality.



Despite the improved quality, the PLMs based news recommenders are limited in multiple perspectives. For example, the recommendation results are lack of explainability \cite{wang2019explainable,liu2020kred}; besides, the deep models are also incompatible with the existing inverted index recall systems \cite{schutze2008introduction,dai2020context}. However, one of the most critical challenges about PLMs based news recommenders is efficiency. While making news recommendations for a user, her entire browsing history needs to be encoded by pre-trained language models. Given the expensive computation costs of PLMs, it will be extremely difficult to make real-time recommendations for those active users who are associated with lots of historical news clicks. In recent years, many approaches are proposed for the acceleration of PLMs. One popular way is to distill large-scale PLMs into lightweight models \cite{sanh2019distilbert,sun2019patient}, where the computation cost can be saved proportionally with the reduction of model size. Besides, efficient transformers are proposed, which may reduce the time complexity of self-attention from quadratic to linear (or log-linear). For example, Linformer \cite{wang2020linformer} and Performer \cite{choromanski2020rethinking} leverage low-rank self-attention; Sparse Transformers \cite{child2019generating} and Big Bird \cite{zaheer2020big} utilize sparse self-attention; Reformer introduces learnable attention patterns, and Synthesizer \cite{tay2021synthesizer} introduces randomized attention patterns. 

However, we argue that the existing acceleration approaches are still not fast enough to handle the challenges in news recommendation. This is because online users are constantly generating large amounts of news clicks. Therefore, even with the proportionally reduced cost from distillation (e.g., by half or one quarter) or the linearly growing cost from efficient transformers, there will still be formidable costs for the recommenders. In this work, we tackle the efficient news recommendation from the perspective of input filtering, which enjoys the merits of reducing the computation cost by multi-fold while still keeping almost the same recommendation quality. Besides, it should also be noted that our work is complementary to the existing acceleration methods, which can be jointly used for further speedup. 

\section{Methodologies}


The news recommendation problem is formulated based on the typical definitions \cite{wu2020mind}. Particularly, given user's historical news clicks $\mathcal{S}=\{S_1,...,S_N\}$, the recommender learns to represent the user's underlying interest, and predicts the user's future news clicks based on it. In fact, the news recommendation problem can be regarded as a special case of document matching: the historical news articles are treated as the query, and the future news articles become the documents to be matched. One unique challenge about news recommendation is that the user-side input can be super large: one user may generate a huge amount of historical news clicks, each of which is an article consisting of multiple sentences (headlines, abstract, body content, etc.) As a result, the cost of encoding the entire user-side input with PLMs will be prohibitive. 

\begin{figure}[t]
\centering
\includegraphics[width=0.99\linewidth]{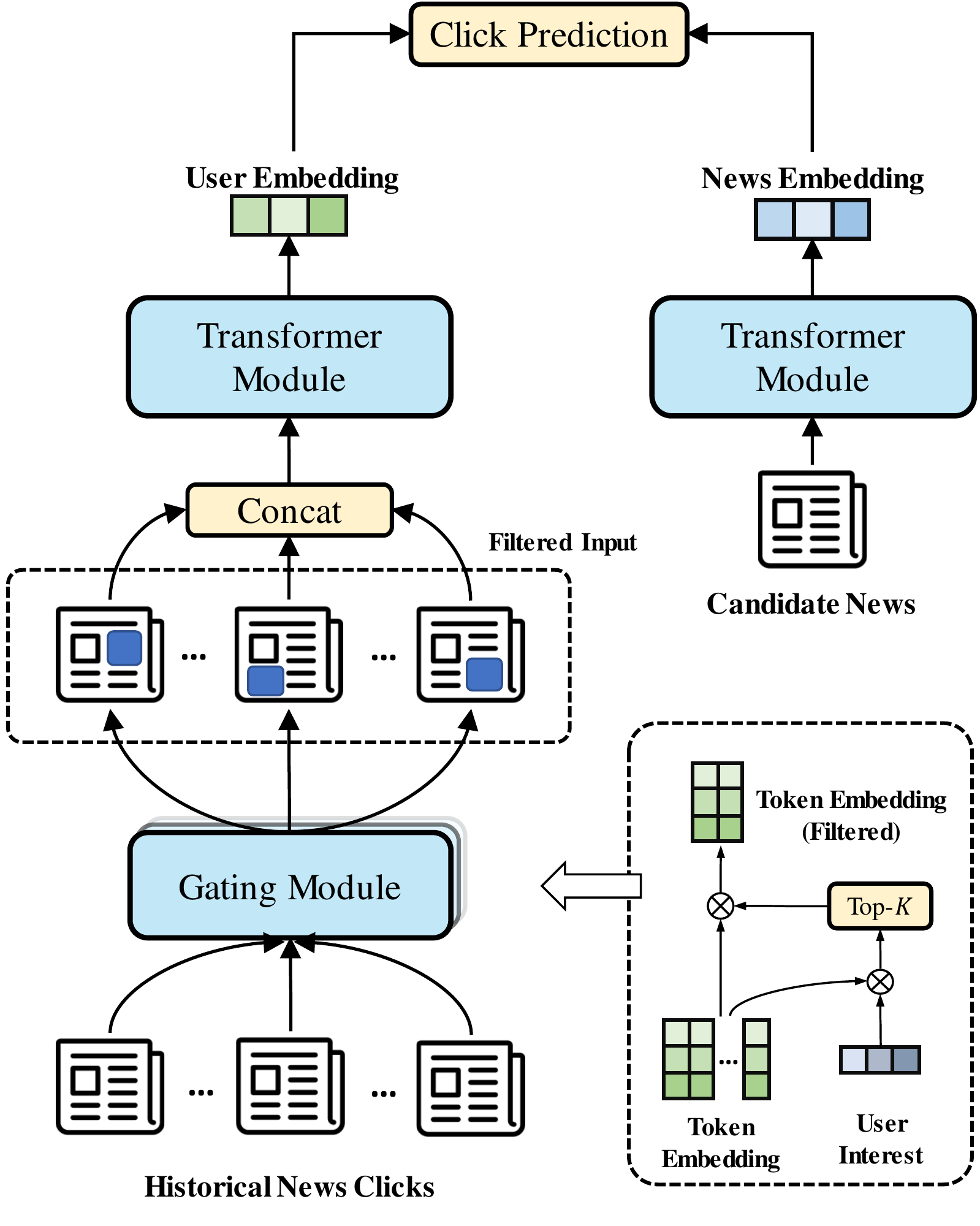}
\caption{\small Overview of GateFormer. The historical news clicks (i.e., the user-side input) are filtered by the gating module, from which the representative keywords are selected for each news article based on user's underlying interests. The filtered input (i.e., the selected keywords) is concatenated and encoded by the transformer module, where the user embedding is produced for the click prediction of candidate news.}
\vspace{-10pt}
\label{fig:1}
\end{figure}

\subsection{Overview of GateFormer}
The framework of GateFormer is shown in Figure \ref{fig:1}. There are two basic components in GateFormer. First, the Gating Module, which processes the entire user-side input and selects the top-$K$ keywords for each historical news article. The filtered input will be much smaller (usually over 10$\times$ smaller than the original size). Second, the Transformer Module, which encodes the filtered input into a user embedding. Finally, the click prediction is made based on the embedding similarity between user and news. (Although a siamese-encoder-like framework is presented here, people may easily change it into a cross-encoder with very little adaptation.) 

\subsection{Gating Module}


$\bullet$ \textbf{User interest encoding}. The gating module performs keyword selection based on the user's interest. For efficiency concerns, the user interest is encoded by the following \textbf{lightweight} models.

Assume that each news article is tokenized into a list of tokens: $S_i=\{t_1,...,t_L\}$ (we inherit the WordPiece tokenizer used by BERT). Firstly, we leverage a single-layer 1D-CNN to generate the context-aware embedding for each input token:
\begin{equation}\label{eq:1}
    \mathbf{h}_j^i = \mathsf{ReLU}(\mathbf{F}\times \mathbf{e}_{j-w:j+w}^i + \mathbf{b}).
\end{equation}
Here, $\mathbf{e}^i_j$ is input word embedding for the $j$-th token in $S_i$; $w$ is the window size; $\mathbf{F}\in\mathbb{R}^{N_f\times(2w+1)d}$ and $\mathbf{b}\in\mathbb{R}^{N_f\times1}$ are the network's parameters ($N_f$ is the number of filters). Then, we aggregate the context-aware embeddings with weighted pooling:
\begin{equation}\label{eq:2}
\begin{aligned}
  & \mathbf{h}_i = \sum \alpha_j * \mathbf{h}_j^i \in \mathbb{R}^{d{\times}1}, \\
  & \alpha_i = \frac{\exp{(\mathbf{v}^T\mathbf{h}_j^i)}}{\sum_{\mathbf{h}_*^i}\exp{(\mathbf{v}^T\mathbf{h}_*^i)}},
\end{aligned}
\end{equation}
where $\mathbf{v}\in\mathbb{R}^{d\times1}$ is the learnable vector for the pooling weights. Finally, we further aggregate the pooling results of all the historical news clicks for the user with a single-layer LSTM:
\begin{equation}\label{eq:3}
    \mathbf{u}^g = \mathsf{LSTM}([\mathbf{h}_1,...,\mathbf{h}_N]).
\end{equation}
Here, $\mathbf{u}^g$ is the aggregation result used for the representation of user interest (by the gating module); we take the last output hidden-state for $\mathbf{u}^g$. 

$\bullet$ \textbf{Keyword selection}. The keywords are selected from the tokens of each news article w.r.t. user's interest. Particularly, the input tokens are selected such that may fully represent the user interest. Therefore, we use the correlation with the extracted user interest as the indicator of a token's importance. In this place, the following cosine similarity is calculated between $\mathbf{u}^g$ and each input token (e.g., the $j$-th token in news $\mathcal{S}_i$):
\begin{equation}\label{eq:4}
    \mathbf{r}_j^i = \mathsf{cos}(\mathbf{u}^g, \mathbf{h}_j^i).
\end{equation}

Next, we make the Top-$K$ selection of keywords for each news based on the importance scores. One notable feature is that the selection is made \textbf{fully differentiable} such that it can be incorporated in an end-to-end training process:
\begin{equation}\label{eq:5}
\begin{aligned}
    &1)~\mathbf{x}^i = \mathsf{argTopK}(\mathbf{r}^i) \in \mathbb{R}^{K\times 1},\\ 
    &2)~\mathbf{y}^i = \mathsf{onehot}(\mathbf{x}^i)\in \mathbb{R}^{K\times L},\\
    &3)~ \mathbf{\widehat{e}}^i = \mathbf{y}^i\mathbf{{e}}^i\in\mathbb{R}^{K\times d}.
\end{aligned}
\end{equation}
In the above equations, we first get the IDs of the tokens ($\mathbf{x}^i$) that produce the Top-$K$ importance scores. Then, we convert IDs into one-hot vectors $\mathbf{y}^i$ (formed as a $K\times L$ matrix). Finally, the embeddings for the selected tokens $\mathbf{\widehat{e}}^i$ ($K\times D$) are extracted from the original token embedding matrix $\mathbf{e}^i$ ($L\times D$) through the multiplication with $\mathbf{y}^i$. Here, $\mathbf{y}^i$ is detached as a constant template matrix while multiplying with $\mathbf{{e}}^i$; thus, 
the whole computation is fully differentiable w.r.t. $\mathbf{e}^i$. We implement a non-duplicated selection by only counting the first appearance of each token and masking the duplicated ones in the template matrix $\mathbf{y}^i$); as a result, there will be no duplications for the selected tokens within each news article. 

It should be noted that there is no explicit supervision signal for the gating module; thus, it has to be learned together with the transformer module based on the news click prediction. In this place, each selected token is weighted by its normalized importance estimated by the gating module:
\begin{equation}\label{eq:6}
\begin{aligned}
    &\mathbf{\widehat{e}}^i = \beta^i \odot \mathbf{\widehat{e}}^i \in \mathbb{R}^{K\times d}, \\
    &\beta^i = \mathsf{softmax}(\mathbf{y}^i\mathbf{r}^i),\,\,
\end{aligned}
\end{equation}
where ``$\odot$'' denotes Hadamard product. By doing so, the gating module becomes cascaded with the transformer module in the resulted computation graph; therefore, it can learn to select the tokens that optimize the news click prediction.


\subsection{Transformer Module}

The filtered input is encoded by the transformer module for user embedding. Firstly, the selected tokens from all the historical news articles are concatenated for transformer's encoding ($\mathsf{Trans}$):
\begin{equation}\label{eq:7}
    \mathbf{g}^\mathrm{u} = \mathsf{Trans}(\mathbf{e}^\mathrm{u}) \in \mathbb{R}^{(N{\times}K){\times}d}.
\end{equation}
In this place, $\mathbf{e}^\mathrm{u}$ denotes the concatenation of the selected token embeddings from the entire user history, i.e., $\mathbf{e}^\mathrm{u} = \bigoplus_{i=1}^N\mathbf{\widehat{e}}^i$. Secondly, the transformer's encoding results are further aggregated for the user's embedding via weighted pooling:
\begin{equation}\label{eq:8}
\begin{aligned}
  & \mathbf{u}^t = \sum \alpha_i * \mathbf{g}_i^\mathrm{u} \in \mathbb{R}^{d{\times}1}, \\
  & \alpha_i = \frac{\exp{(\mathbf{q}^T\mathbf{g}_i^\mathrm{u})}}{\sum_{\mathbf{g}_*^\mathrm{u}}\exp{(\mathbf{q}^T\mathbf{g}_*^\mathrm{u})}},
\end{aligned}
\end{equation}
where $\mathbf{q} \in \mathbb{R}^{d{\times}1}$ is the learnable vector for the pooling weights. Similarly, the candidate news embedding is generated as follows:
\begin{equation}\label{eq:9}
    \mathbf{c}^t = \mathsf{WPA}(\mathsf{Trans}(\mathbf{e}^\mathrm{c}), \mathbf{q}).
\end{equation}
In this place, $\mathbf{e}^\mathrm{c}$ is the concatenation of token embeddings for the candidate news: $\mathbf{e}^\mathrm{c} = \bigoplus_{i=1}^L \mathbf{e}^\mathrm{c}_i$; $\mathsf{WPA}(\cdot)$ stands for the weighted pooling based aggregation in Eq. \ref{eq:8}. Here, the transformer module's parameters are shared for the encoding process of user and candidate news. 
The click prediction is made based on the similarity between user ($u$) and candidate news ($c$). Here, we choose the scaled inner product for the measurement of similarity:
\begin{equation}
    z(u,c) = {\langle \mathbf{u}^t, \mathbf{c}^t \rangle}/{\sqrt{d}}.
\end{equation}
The model is learned to predict the user's future news clicks ($\mathcal{C}^+$) in contrast to the negative samples. Following the common practice on news recommendation \cite{wu2020mind,liu2020kred}, we take the impressed but non-clicked news articles as our negative samples ($\mathcal{C}^-$). Finally, the loss function is formulated for each user as below:
\begin{equation}
    \mathcal{L}^u = -\sum_{c\in \mathcal{C^+}}\log\frac{\exp(z(u,c))}{\sum\limits_{c'\in\{\mathcal{C}^+,\mathcal{C}^{-}\}}\exp({z}(u,c'))}.
\end{equation}
The loss function is minimized for all the users so as to optimize the model's click prediction. 

$\bullet$ \textbf{Efficiency Analysis}. Let $\mathcal{I}_{org}$ and $\mathcal{I}_{flt}$ denote the original and filtered user input, respectively. The overall time cost for GateFormer can be derived as: $\mathrm{T}_{gate}(\mathcal{I}_{org})+\mathrm{T}_{trans}(\mathcal{I}_{flt})$. Given that the gating module's time cost is linearly growing, we have: $\mathrm{T}_{gate}(\mathcal{I}_{org}) \propto \lambda_1|\mathcal{I}_{org}|$, where $\lambda_1$ is the gating module's unit time cost to process one single token. Besides, because the transformer module's time cost is polynomially growing (the self attention is quadratic, other computations are linear), we have: $\mathrm{T}_{trans}(\mathcal{I}_{flt}) > \lambda_2|I_{flt}|$, where $\lambda_2$ is transformer's unit time cost to process one single token. As a result, we may derive the following acceleration ratio for GateFormer:
\begin{equation}
\begin{aligned}
    \gamma 
    &= \frac{\mathrm{T}_{trans}(\mathcal{I}_{org})}{\mathrm{T}_{gate}(\mathcal{I}_{org})+\mathrm{T}_{trans}(\mathcal{I}_{flt})}, \\
    &> \frac{\lambda_2|\mathcal{I}_{org}|}{\lambda_1|\mathcal{I}_{org}|+\lambda_2|\mathcal{I}_{flt}|}, \\
    &= \frac{1}{{\lambda_1}/{\lambda_2}+{|\mathcal{I}_{flt}|}/{|\mathcal{I}_{org}|}}.
\end{aligned}
\end{equation}
Note that $\lambda_2 \gg \lambda_1$ even with a distilled BERT like PLM used for the transformer module, $\gamma$ will be almost $|\mathcal{I}_{flt}|/|\mathcal{I}_{org}|$ in reality. In our experiment, we demonstrate that GateFormer achieves on-par performance as SOTA with more than 10-fold compression of input, which means over $10\times$ lossless acceleration of the news recommender.

\begin{table*}[t]
    \centering
    \small
    \caption{\small Evaluations of accuracy and efficiency on MIND. The highest performances are bolded, the strongest baselines (GFMs excluded) are underlined. (Time: time cost per user's inference; Mem: GPU RAM usage with $\textit{batch}\_\textit{size}$=32.)}
    \vspace{-5pt}
    \label{tab:effectiveness_mind} 
    \begin{tabular}{p{0.10\linewidth}p{0.12\linewidth}cccc|ccc}
    \toprule
    
    & & \multicolumn{4}{c}{Effectiveness} & \multicolumn{3}{c}{Efficiency}\\
    \cmidrule(lr){3-6}
    \cmidrule(lr){7-9}
    
    \textbf{Type} & \textbf{Methods} & \textbf{AUC} & \textbf{MRR} & \textbf{NDCG@5} & \textbf{NDCG@10} & \textbf{Time}$/\mathrm{ms}$& \textbf{Mem}$/\mathrm{MB}$ & \textbf{FLOPs}$/\mathrm{G}$\\
    \midrule

    \multirow{4}{0.12\linewidth}{Standard Methods} 
    & NAML & $66.86$ & $32.49$ & $35.24$ & $40.19$ & $0.09$ & $2490$ & $0.4$\\
    & LSTUR & $67.73$ & $32.77$ & $35.59$ & $41.34$ & $0.13$ & ${6942}$ & $0.5$\\
    & NRMS & $67.76$ & $33.05$ & $35.94$ & $41.63$ & $0.16$ & ${2384}$ & $0.7$\\
    & {EBNR} & $\underline{70.42}$ & $\underline{35.07}$ & $\underline{38.40}$ & $\underline{44.12}$ & $5.27$ & $6580$ &  $193.4$\\
    \midrule
    \multirow{6}{0.12\linewidth}{Efficient Trans. $\&$ Distilled-PLMs}
    & BigBird & $68.41$ & $33.82$ & $36.55$ & $42.26$ & $7.02$ & $5766$ & ${130.5}$\\
    & LongFormer & $68.02$ & $33.07$ & $36.14$ & $41.72$ & $6.33$ & ${5536}$ & $130.5$\\
    & Synthesizer & $68.49$ & $33.90$ & $36.87$ & $42.54$ & $3.79$ & $5964$ & $193.4$\\
    & Funnel Trans. & $69.88$ & $34.13$ & $37.29$ & $43.09$ & $4.96$ & $8140$ & $161.2$\\
    & DistilBert & $70.39$ & $34.97$ & $38.29$ & $44.01$ & $2.70$ & $5060$ & $97.1$\\
    & NewsBert & $70.31$ & $34.89$ & $38.32$ & $43.95$ & $1.85$ & $4336$ & $64.8$\\
    \midrule
    
    \multirow{4}{0.12\linewidth}{Variational Filters} 
    & GFM (First) & $70.08$ & $34.83$ & $38.12$ & $43.83$ & $0.66$ & $2658$ & $20.0$\\
    & GFM (BM25) & $69.36$ & $34.28$ & $37.44$ & $43.17$ & -- & -- & -- \\
    & GFM (Entity) & $69.76$ & $34.47$ & $37.67$ & $43.41$ & -- & -- & -- \\
    & GFM (Key) & $69.81$ & $34.49$ & $37.71$ & $43.50$ & -- & -- & -- \\
    
    \midrule
    
    \multirow{2}{0.12\linewidth}{Variational Gates}
    & GFM (Trans) & $70.58$ & $35.12$ & $38.46$ & $44.18$ & $1.00$ & $3660$ & $25.4$\\
    & GFM (ATTN) & $70.21$ & $34.96$ & $38.32$ & $43.95$ & $0.74$ & $3244$ & $20.2$ \\ 

    \midrule
    \multirow{2}{0.12\linewidth}{Variational Schemes}
    & GFM (Global) & ${70.48}$ & ${35.14}$ & ${38.53}$ & ${44.22}$ & ${0.67}$ & ${3158}$ & ${20.4}$\\
    & GFM (TTW) & ${70.40}$ & ${34.65}$ & ${37.89}$ & ${43.68}$ & ${0.89}$ & ${3202}$ & ${20.0}$\\
    \midrule
    \multirow{1}{0.12\linewidth}{Ours}
    & GateFormer & $\mathbf{70.97}$ & $\mathbf{35.50}$ & $\mathbf{38.93}$ & $\mathbf{44.63}$ & $\mathbf{0.79}$ & $\mathbf{3376}$ & $\mathbf{20.5}$\\
    
    \bottomrule
  \end{tabular}   
  \vspace{-10pt}
\end{table*}

\section{Experiments}
Our experiments are mainly dedicated to study GateFormer's impact on news recommendation accuracy and efficiency. Besides, we explore its application to keyword based news retrieval, demonstrate its interpretability, and evaluate its effectiveness on other corpus. 

$\bullet$ \textbf{Data}. Our experimental studies are mainly based on \textbf{MIND} dataset~\cite{wu2020mind}, which is the largest open benchmark on news recommendation with rich textual features. This dataset includes 1,000,000 users, 161,013 news articles, and 24,155,470 news clicks from Microsoft News. We follow the standard settings of this dataset, where there are 2,186,683 samples in the training set, 365,200 samples in the validation set, and 2,341,619 samples in the test set. The title and abstract of each news article are concatenated as its input feature.
Besides, we also leverage another large-scale dataset  \textbf{DBLP}\footnote{\scriptsize  https://originalstatic.aminer.cn/misc/dblp.v12.7z} to explore our effectiveness in applications beyond news recommendation. This dataset contains academic papers and their citation relationships from the latest dump of DBLP. There are 4,894,081 papers in total; each paper is associated with its title and, has 9.31 references on average. The \textit{reference recommendation} task is evaluated by: given a paper and its citations (prediction target omitted), the model is learned to predict whether another paper will also be included in the reference list.

$\bullet$ \textbf{Baselines}. The following classes of baseline methods are utilized for our experiments. 

Firstly, we compare with the following \textit{Standard Methods}. 1)
\textbf{NAML} \cite{wu2019neural}, 2) \textbf{LSTUR} \cite{an2019neural},
3) \textbf{NRMS}~\cite{wu2019nrms}, which uses multi-layers of self-attention networks for news and user encoding; it is reported to be the strongest non-PLMs baseline in~\cite{wu2020mind}. The above methods purely rely on lightweight networking components; thus, the computation costs are relatively smaller than the PLMs based recommenders. we include 4) \textbf{EBNR}~\cite{wu_newsPLM} as the representative for the standard \textit{PLMs-based news recommenders}. The proposed method outperforms typical options like BERT and RoBERTa, and achieves SOTA single model performance on news recommendation. 

Secondly, we make use of the following \textit{accelerated PLMs-based} methods. One is based on \textit{efficient transformers}, including 5) \textbf{Big Bird}~\cite{zaheer2020big}, 6) \textbf{LongFormer}~\cite{beltagy2020longformer}, 7) \textbf{Synthesizer}~\cite{tay2021synthesizer}, and 8) \textbf{Funnel Transformers}~\cite{dai2020funnel}. The other one is based on \textit{distilled lightweight PLMs}, including a widely used distillation model in general: 9) \textbf{DistillBERT}~\cite{sanh2019distilbert}, and a recently proposed distillation model on Microsoft News corpus: 10) \textbf{NewsBERT}~\cite{wu2021newsbert}. 

Thirdly, we introduce variations of GateFormer (GFM) by filtering input with the following methods. 11) \textbf{GFM (BM25)}, where the user-side input is filtered by BM25 score; 12) \textbf{GFM (Entity)}, where the entities within each news article are filtered; 13) \textbf{GFM (First-$K$)}, where the first $K$ tokens are kept; and 14) \textbf{GFM (KeyBERT)}: where KeyBERT~\cite{keybert} (a pre-trained model for keyword extraction) is used to generate the filtered input.

Finally, we test different networks for the gating module of GateFormer. 14) The news encoder in gate is switched to transformer: \textbf{GFM (Trans)}; and 15) the user encoder in gate is switched to attention pooling: \textbf{GFM (ATTN)}. 

$\bullet$ \textbf{Evaluation Metrics}. We evaluate the recommendation's accuracy based on typical ranking metrics, like \textit{AUC}, \textit{NDCG}, \textit{MRR}, etc. Besides, we also evaluate the efficiency in terms of \textit{Inference Time}, \textit{Memory Consumption}, and \textit{FLOPs}. 

Following the default settings in \cite{wu2020mind}, the user is uniformly truncated to 50 historical news clicks. The news article is truncated to \textbf{30} tokens. For GateFormer and other GFM variations, the \textbf{Top-$3$} tokens are selected by default. That is to say, the input size is compressed by \textbf{10-fold}. More compression ratios will also be evaluated in our extended analysis. Our implementation is based on PyTorch-1.9.1. The experiments are performed on a cluster of 2* Nvidia A100 (40GB) GPUs, and 2* AMD EPYC 7v12 64-core CPUs. \textit{Supplementary experiments, configurations about training/testing are specified in Appendix. Our code will be open-sourced after the review stage. }

\begin{table*}[t]
  \centering
  \small
  \caption{\small News Recall Evaluations.}
  \vspace{-5pt}
  \label{table:recall}
  \begin{tabular}{llccc}
    \toprule
    \textbf{Type} & \textbf{Methods} & \textbf{Recall@10} & \textbf{Recall@50} & \textbf{Recall@100}\\
    \midrule
    \multirow{2}{0.12\linewidth}{Dense} 
    & EBNR & $5.01$ & $16.37$ & $27.71$\\
    & GateFormer (D) & $6.31$ & $18.57$ & $29.13$\\
    \midrule
    \multirow{3}{0.12\linewidth}{Sparse}
    & BM25 & $1.46$ & $5.00$ & $8.43$\\
    & KeyBERT & $2.35$ & $10.42$ & $16.35$\\
    & GateFormer (S) & $2.84$ & $10.64$ & $18.10$\\
    \midrule
    \multirow{3}{0.12\linewidth}{Sparse+Dense}
    & BM25+EBNR & $2.89$ & $8.92$ & $13.68$\\
    & KeyBERT+EBNR & $5.57$ & $14.66$ & $21.81$\\
    & GateFormer (S+D) & $6.39$ & $18.23$ & $28.74$ \\
    \bottomrule
  \end{tabular}
  \vspace{-10pt}
\end{table*}

\subsection{Experiment Analysis}
First of all, we evaluate the recommendation accuracy and efficiency of different methods on MIND dataset, whose results are shown in Table \ref{tab:effectiveness_mind}. 

$\bullet$ \textit{Comparison with the Standard.} We may find that the methods based on lightweight networks: \textit{NAML}, \textit{LSTUR}, \textit{NRMS}, are far more efficient than other methods leveraging multi-layer transformers or PLMs. However, such lightweight recommenders are highly limited in recommendation accuracy: the resulted performances are the worst of all among all the comparison methods. The other extreme of the standard methods is EBNR. On one hand, it is extremely slow: the average running time and FLOPs are $\times10$s or even $\times100$s greater than the lightweight recommenders; on the other hand, the recommendation accuracy can be notably improved on top of the usage of PLMs. 

Compared with the above standard methods, GateFormer is highly competitive on both ends. Firstly, it leads to $\times6.7$ speedup on inference time and $\times9.4$ on FLOPs over EBNR. Such observations are consistent with the fact that the user-side input is compressed by 10-fold in GateFormer. Secondly, it achieves the highest recommendation accuracy among all the approaches in comparison (even slightly outperforms EBNR). The competitiveness in accuracy can be explained as follows. On one hand, GateFormer effectively preserves the informative keywords in user history, which may fully capture users' underlying interests. On the other hand, given that the user-side input is significantly simplified, merely consisting of less than 150 tokens (the top-3 keywords from 50 historical news clicks), we are able to concatenate all of them for transformer's encoding. By doing so, each input token becomes ``fully context-aware'': it may not only refer to the tokens within the same news article, but also get aware of the information within the entire user history. As a result, we may derive an in-depth understanding of user interest from the encoding result. In contrast, conventional PLMs based methods, like EBNR, cannot jointly model the entire news clicks in user history given the restriction on input length (usually 512 tokens); thus, they have to encode each news click individually, which is inferior to our fully context-aware encoding result.

$\bullet$ \textit{Comparison with Accelerated PLMs}. We further analyze two classes of methods that may accelerate the PLMs based recommenders. 

Firstly, we study efficient transformers, including \textit{Big Bird}, \textit{LongFormer}, \textit{Synthesizer}, and \textit{Funnel Transformer}. The first three methods reduce the time complexity of self attention, while the last approach performs a layer-wise reduction of the hidden states. Despite the theoretical acceleration, we find the cost reduction is very limited in reality; for Big Bird and LongFormer, the running time is even higher than EBNR, which uses vanilla BERT (base) like PLMs (in fact, this is consistent with the observations in Long Range Arena \cite{tay2020long}). As discussed, the self-attention only accounts for a limited portion of PLMs' overall computation cost (especially when the input sequence merely consists of tens or hundreds of tokens). Therefore, the cost reduction of self-attention will not substantially resolve the bottleneck on efficiency. For Funnel transformer: it still has to preserve a large fraction of hidden-states so as to maintain the representation quality; thus, the overall time reduction is marginal as well. Besides the limited effect on efficiency, all the above approaches suffer from severe loss of recommendation accuracy compared with EBNR. 

Secondly, we analyse the distillation based methods, including \textit{DistillBERT} and \textit{NewsBERT}. It can be observed that both methods achieve competitive recommendation accuracy, which is close to EBNR's performance. Besides, the recommendation efficiency is notably improved as well, given that the distilled models' scales are merely half or one-third of BERT base. Overall, the distillation based methods are more appropriate for the acceleration of news recommendation. However, it should also be noted that these methods are still inferior to GateFormer in terms of both accuracy and efficiency. 

$\bullet$ \textit{Comparison with GFM variations}. We further compare four variations of GateFormer: \textit{GFM (First)}, \textit{GFM (BM25)}, \textit{GFM (Entity)}, and \textit{GFM (Key)}. All these approaches rely on pre-defined heuristics for input filtering. Compared with GateFormer, the recommendation efficiency can be slightly improved, knowing that the time cost is saved from the gating module (we only report the computation cost for \textit{GFM (First)}, as all of them enjoy the same working efficiency). However, it may also be observed that the reduction of time cost is relatively small, which reflects that our gating module is efficient in reality. Besides, we find that GateFormer achieves notably higher recommendation accuracy compared with all the GFM variations, which indicates that our gating module is effectively learned to select informative keywords from user-side input. 

\begin{figure}[t]
    \centering
    \includegraphics[width=0.92\linewidth]{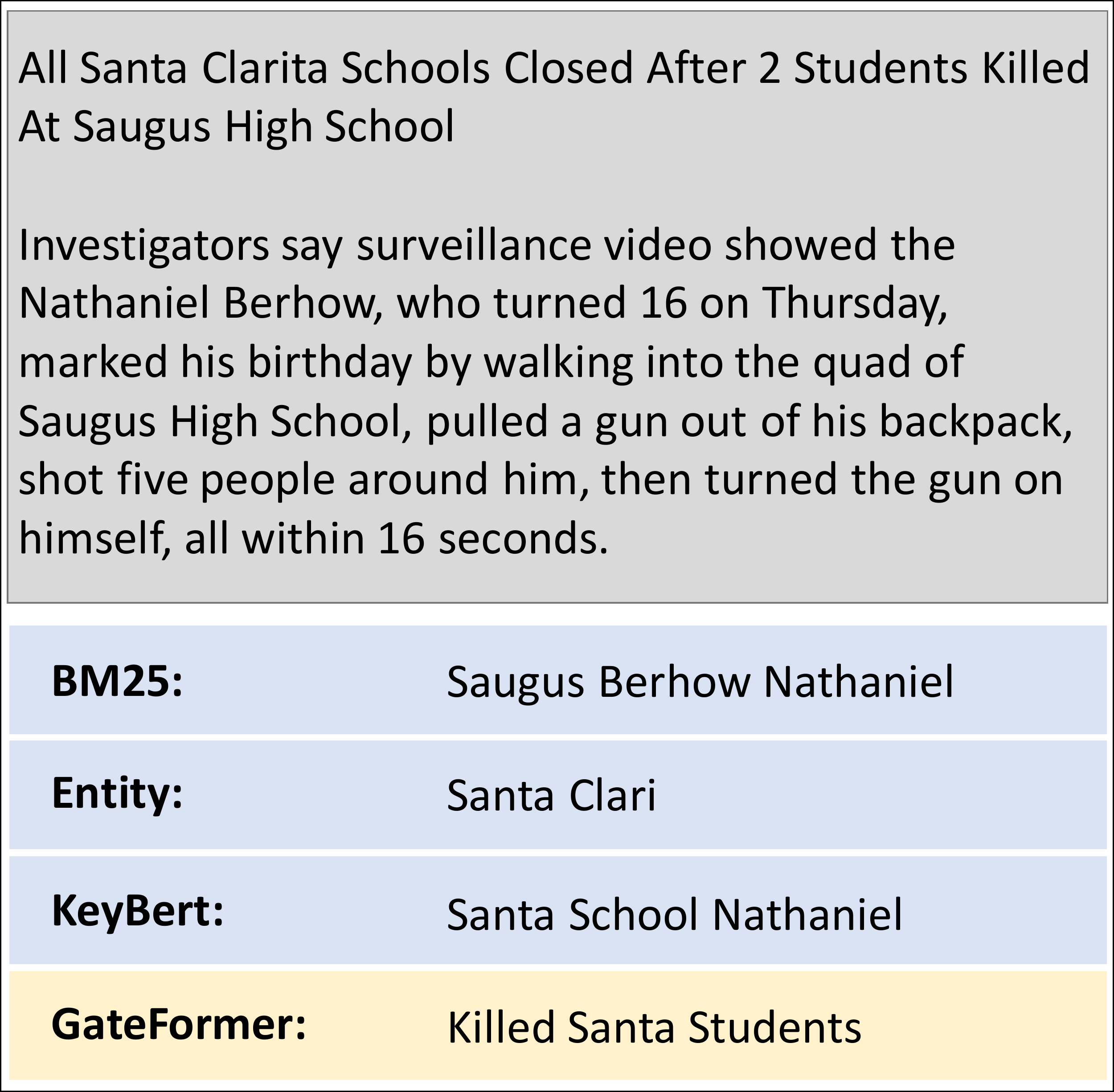}
    \caption{\small Case analysis. The upper box shows the content of the original news article, the lower boxes show the selected keywords from different methods.}
    \vspace{-10pt}
    \label{fig:case}
    \vspace{-10pt}
\end{figure}

$\bullet$ \textit{Comparison with Gate Variations}. We analyze GateFormer's performances with two variational gates: \textit{GFM (Trans)} uses a more complicated news encoder, while \textit{GFM (ATTN)} uses a simplified user encoder. Despite the slight differences in time cost, none of them is able to improve the recommendation accuracy of the default GateFormer. So far, the CNN+LSTM combination remains the best configuration for the gating module. Note that it is inappropriate to use more complicated networks, as the gating module needs to process the entire user input. In our experiments, the cost from the gating module will become formidable even with a 2-layer transformer. 

$\bullet$ \textit{News recall evaluation}. We evaluate the GateFormer's performance for news recall in Table \ref{table:recall}. Firstly, analyze the dense recall scenario, where the candidate news are recalled based on the similarity between the user and news embeddings. We find that GateFormer achieves higher recall rates than EBNR, which is consistent with the ranking performance reported in Table \ref{tab:effectiveness_mind}. Secondly, we analyze the sparse recall scenario, where the candidate news is recalled based on the identified keywords. We find that GateFormer outperforms the baseline methods with notable advantages. We also analyze the effect of sparse and dense retrieval: the candidate news are firstly retrieved by keywords, and then get re-ranked for the finalist based on the embedding similarity. Once again, GateFormer substantially improves the recall rates against the baselines in comparison. Such observations indicate that GateFormer may serve as a strong enhancement of the current keyword selection mechanisms, and get seamlessly integrated into the existing inverted index systems. 

\begin{table}[t]
    \centering
    \small
    \caption{\small Evaluation on DBLP.}
    \vspace{-5pt}
    \label{tab:effectiveness_dblp}
    \begin{tabular}{lccccc}
    \toprule
    \textbf{Methods} & \textbf{P@1} & \textbf{P@5} & \textbf{MRR} & \textbf{NDCG}\\
    \midrule
    GFM (Random) & $40.23$ & $66.82$ & $52.50$ & $62.36$ \\
    GFM (First) & $50.73$ & $75.30$ & $61.21$ & $69.51$\\
    GFM (BM25) & $43.89$ & $69.48$ & $55.70$ & $64.93$\\
    GFM (KeyBERT) & $46.46$ & $72.51$ & $58.38$ & $67.17$ \\
    \midrule
    GateFormer & $\textbf{64.86}$ & $\textbf{89.03}$ & $\textbf{75.41}$ & $\textbf{81.06}$\\
    \bottomrule
  \end{tabular}
  \vspace{-10pt}
\end{table}

$\bullet$ \textit{Case Analysis}. We introduce a case study to demonstrate the keyword selection from different methods. As shown in Figure \ref{fig:case}, \textit{BM25} tends to select the words which are \textit{unique} to the news article; however, such words do not necessarily represent the underlying semantics about the news article. The \textit{Entity} may highlight some of the informative contents; unfortunately, it is prone to poor coverage of the overall semantics. The \textit{KeyBERT} is able to capture some of the keywords which are informative in general, but probably less important to reflect the news semantics and user interest. By comparison, GateFormer gives rise to the best keywords selection result, which fully represents the news semantics and indicates the underlying reason why it catches the user's attention. 

    

$\bullet$ \textit{Evaluation on DBLP}. We make comparison with other gating methods: \textit{GFM (Random)} (due to the limited entity coverage, the entity based method is switched to random selection), \textit{GFM (First)}, \textit{GFM (BM25)} and \textit{GFM (KeyBERT)}. The experiment results in Table \ref{tab:effectiveness_dblp} are consistent with our previous observations in Table \ref{tab:effectiveness_mind}, and GateFormer's advantages are even highlighted. Such a finding suggests that GateFormer may stand as a generic input filtering method, which facilitates the speedup and keyword selection for text oriented recommendation/retrieval tasks. 

\section{Conclusion}
In this paper, we proposed GateFormer for the speedup of news recommendations. It introduced a gating mechanism, where informative keywords can be filtered from historical news clicks to represent the user's interest. The gating module was designed to be personalized, lightweight, and end-to-end learnable, such that the input filtering can be accurately conducted with little running costs. Our experimental studies verified the effectiveness of GateFormer: it achieved notable improvements in both recommendation accuracy and efficiency against the existing acceleration methods. 

\newpage

\newpage
\bibliography{anthology,acl_zhengliu,tkde,custom}
\bibliographystyle{acl_natbib}

\clearpage

\appendix
\section{Training Settings}
We adopt BERT-base-uncased~\cite{Devlin2019BERT} as our PLM backbone. We set word embeddings' dimension $d$ to $768$. The convolution kernel in the gate has $150$ channels ($N_f$); the window size is set to $3$ ($w$); we use two-side padding of length $1$ and a default stride $1$. The output dimension of LSTM is $150$. 
On MIND dataset, we sample $K=4$ negative instances(which is default settings in related literature), and set the batch size to $64$($32$ per node). 
On DBLP dataset, we adopt in-batch negative sampling of ratio $K=59$, and set the batch size to $240$ ($60$ per node).
In all our experiments, we use Adam~\cite{diederik_Adam} optimizer with a linear scheduler with 10000-step warm up of learning rate $6e^{-6}$. 

\section{Supplementary Analysis}
\begin{figure}[t]
    \centering
    \includegraphics[width=\linewidth]{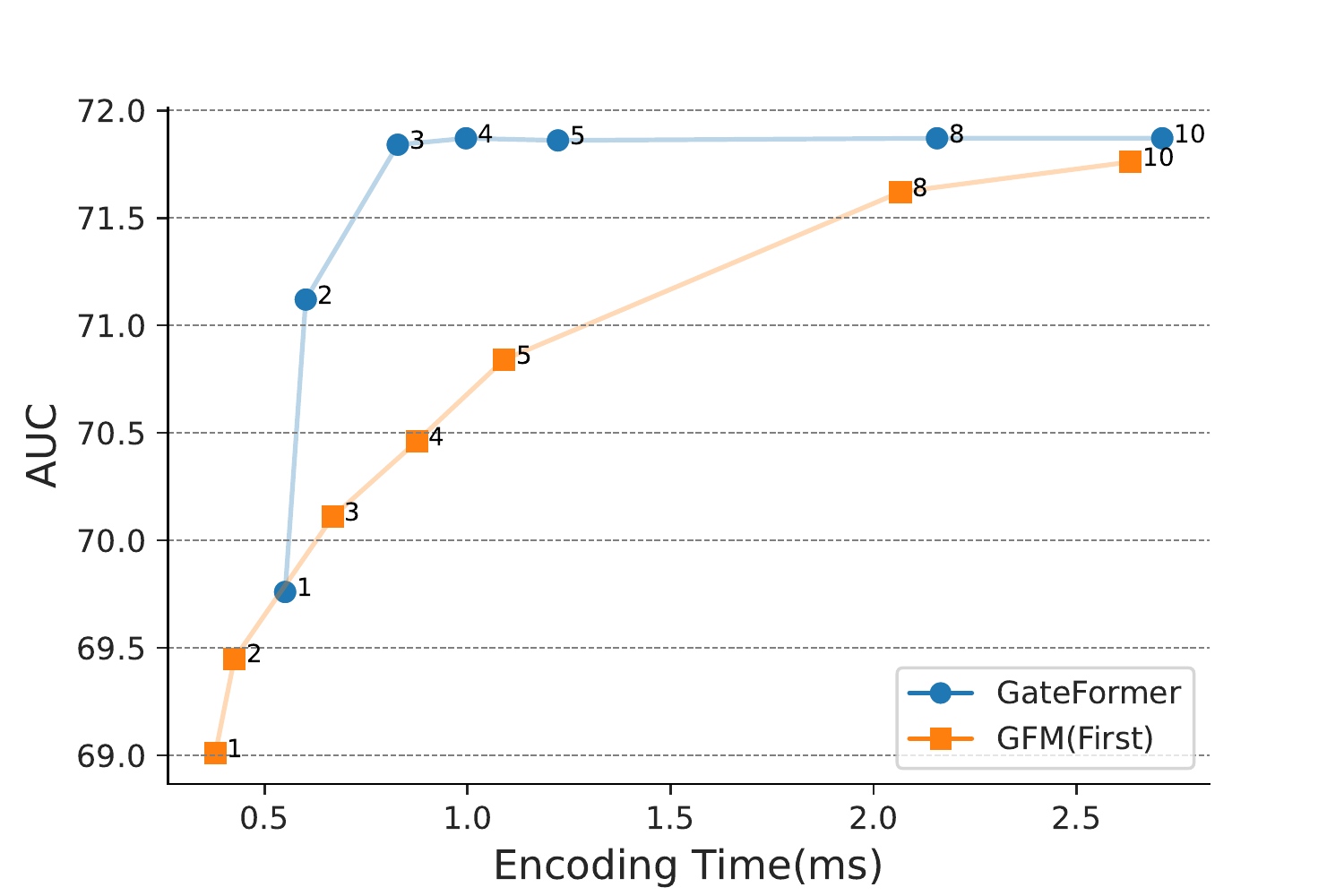}
    \caption{Recommendation accuracy (y-axes) v.s. efficiency (x-axes) with different number of keywords selected per news ($K$). The number next to the marker indicates the value of $K$.} 
    \label{fig:k}
\end{figure}
\begin{figure}[t]
    \centering
    \includegraphics[width=\linewidth]{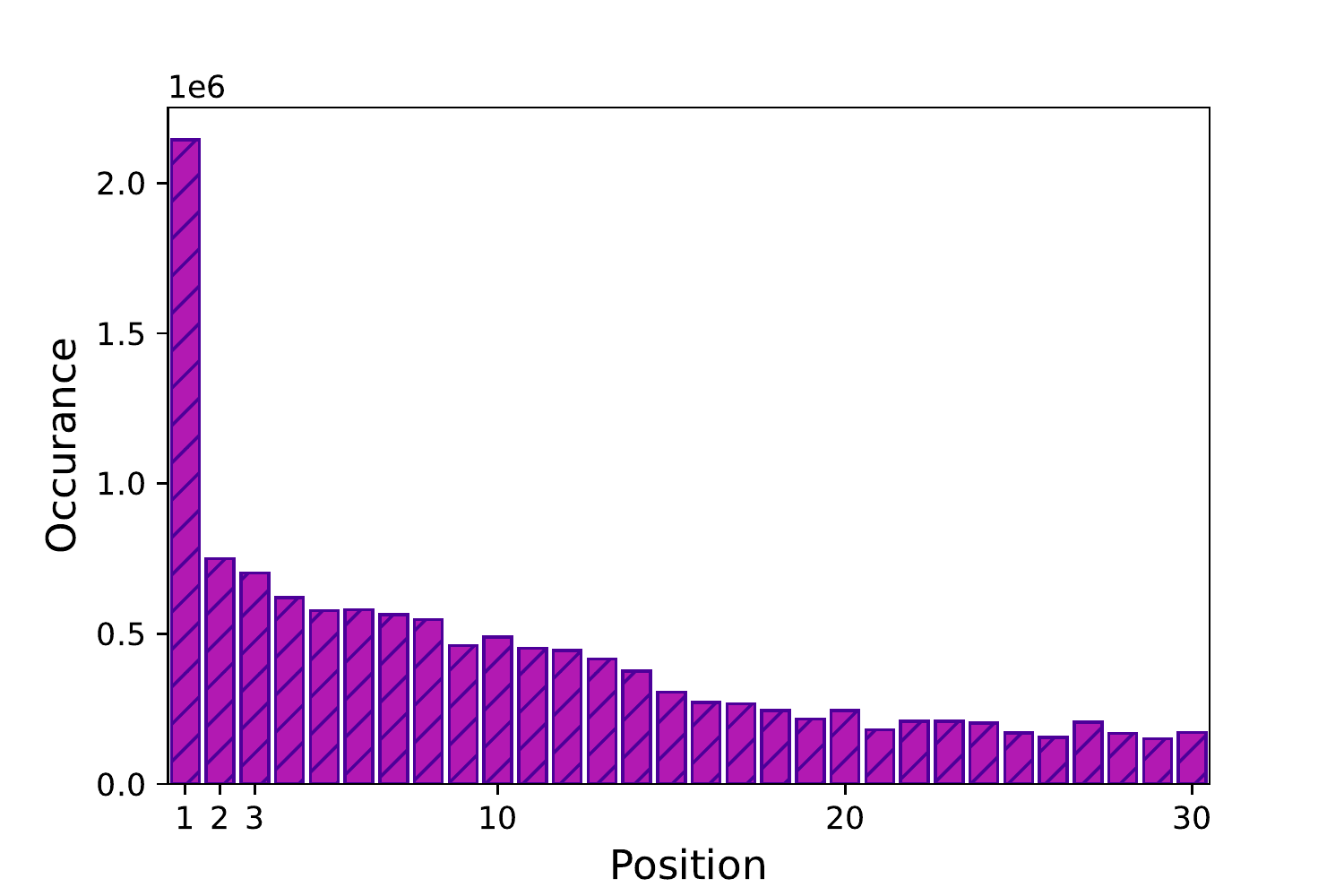}
    \caption{Statistics of the selected keywords' positions in news articles.}
    \label{fig:position distribution}
\end{figure}

\label{subsec:k}
We extend our study by investigating the influence of selecting a different number of tokens per news ($K$). The resulted efficiency and accuracy are visualized in Figure~\ref{fig:k}, where GateFormer is compared against \textit{GFM (First)} (the strongest baseline in Table \ref{tab:effectiveness_mind}). We find that GateFormer is able to consistently outperform GFM {First}, which is consistent with our observation in the default setting. For GateFormer, the performance soon becomes competitive with a small number of tokens selected; while for GFM (First), the accuracy growth is relatively slow. (It is quite interesting to see that three tokens ($K=3$) seem already enough to preserve the underlying semantics of a news article. However, it is should also be noticed that keywords selected in other news articles may complement the information loss in one news article; thus, it's more appropriate to say: \textit{three tokens are almost enough to represent the news semantics within the whole context of user history.})

The final question for GateFormer is: \textit{is keyword selection position relevant?} To answer this question, we visualize the original positions for the selected tokens in Figure~\ref{fig:position distribution}. Overall, the probability of being selected decreases along the position, which means the model tends to select tokens in the front of the news articles (it's quite impressive that the position information about the tokens is not released to GateFormer; however, GateFormer learns to select the tokens which are more likely to appear in the front). Such an observation is consistent with our previous observation that GFM (First) is a relatively competitive baseline; it is also quite intuitive as news articles usually highlight their major points in the very beginning. However, we notice that GateFormer is able to select keywords from the other positions of a news article (because of the long-tail property of the curve, the overall possibility is quite large actually). Such keywords can be equally important and may provide complementary information to the keywords in the front, which contribute to the comprehensive coverage of news semantics.

\end{document}